\documentclass[sigconf]{acmart}

\usepackage{booktabs} 
\usepackage{caption}
\usepackage{graphicx}
\usepackage[ruled,linesnumbered]{algorithm2e}
\usepackage{amsmath,amssymb,amsfonts}
\usepackage{algorithmic}
\usepackage{graphicx}
\usepackage{textcomp}
\usepackage{xcolor}
\usepackage{balance}
\usepackage{subfigure}
\setcopyright{rightsretained}

\acmDOI{10.475/123_4}

\acmISBN{123-4567-24-567/08/06}

\acmConference[WOODSTOCK'97]{ACM Woodstock conference}{July 1997}{El
	Paso, Texas USA} 
\acmYear{1997}
\copyrightyear{2016}

\acmArticle{4}
\acmPrice{15.00}

\editor{Jennifer B. Sartor}
\editor{Theo D'Hondt}
\editor{Wolfgang De Meuter}

\begin{document}
	\title{Multi-dimensional Graph Convolutional Networks}

	\author{Yao Ma}
	\affiliation{%
	  \institution{Michigan State University}
	}
	\email{mayao4@msu.edu}
	
	\author{Suhang Wang}
	\affiliation{%
	  \institution{Arizona State University}
	}
	\email{suhang.wang@asu.edu}
	
	\author{Charu C. Aggarwal}
	\affiliation{%
	  \institution{IBM T. J. Watson Research Center}
}
	\email{charu@us.ibm.com}
	
	\author{Dawei Yin}
	\affiliation{%
	  \institution{JD.com}
	}
	\email{yindawei@acm.org}
	
	\author{Jiliang Tang} 
	\affiliation{%
	 \institution{Michigan State University}
 }
\email{tangjili@msu.edu}
	%
	%
	%
	
	\renewcommand{\shortauthors}{B. Trovato et al.}

\begin{abstract}
Convolutional neural networks (CNNs) leverage the great power in representation learning on regular grid data such as image and video. Recently, increasing attention has been paid on generalizing CNNs to graph or network data which is highly irregular. Some focus on graph-level representation learning while others aim to learn node-level representations. These methods have been shown to boost the performance of many graph-level tasks such as graph classification and node-level tasks such as node classification. Most of these methods have been designed for single-dimensional graphs where a pair of nodes can only be connected by one type of relation. However, many real-world graphs have multiple types of relations and they can be naturally modeled as multi-dimensional graphs with each type of relation as a dimension. Multi-dimensional graphs bring about richer interactions between dimensions, which poses tremendous challenges to the graph convolutional neural networks designed for single-dimensional graphs. In this paper, we study the problem of graph convolutional networks for multi-dimensional graphs and  propose a multi-dimensional convolutional neural network model mGCN aiming to capture rich information in learning node-level representations for multi-dimensional graphs. Comprehensive experiments on real-world multi-dimensional graphs demonstrate the effectiveness of the proposed framework.
\end{abstract}
	
	%
	%



	\maketitle
	\section{introduction}\label{sec:introduction}
Convolutional Neural Networks (CNNs)~\cite{lecun-etal1998gradient} have been proven to bring breakthrough improvements on various tasks on regular grid data such as image, video and speech~\cite{krizhevsky-etal2012imagenet,lawrence-etal1997face,kalchbrenner2014convolutional,kim2014convolutional,karpathy2014large,lecun1995convolutional,abdel2012applying}. However, graph data such as social networks and gene data are highly irregular.  The main issue is that image data  have an extremely high level of spatial locality, which might not be the case in graphs data. 
Recent efforts~\cite{bruna-etal2013spectral,henaff-etal2015deep,duvenaud-etal2015convolutional,li-etal2015gated,defferrard-etal2016convolutional,kipf-etal2016,hamilton-etal2017inductive,schlichtkrull-etal2017modeling} have been made to generalize convolutional neural networks to irregular graph data. Some~\cite{bruna-etal2013spectral,henaff-etal2015deep,duvenaud-etal2015convolutional,li-etal2015gated,defferrard-etal2016convolutional} focus on graph-level representation learning, while others~\cite{kipf-etal2016,hamilton-etal2017inductive,schlichtkrull-etal2017modeling} target on node-level representation learning. These methods have been shown to advance many graph-level tasks such as graph classification~\cite{defferrard-etal2016convolutional} and node-level tasks such as node classification~\cite{kipf-etal2016,hamilton-etal2017inductive,schlichtkrull-etal2017modeling} and link prediction~\cite{schlichtkrull-etal2017modeling}. In this work, we focus on node-level representation learning with graph covolutional neural networks. 


\begin{figure*}
\begin{center}
\subfigure[Single-dimensional graph]{\label{fig:single-relation}\includegraphics[scale=0.55]{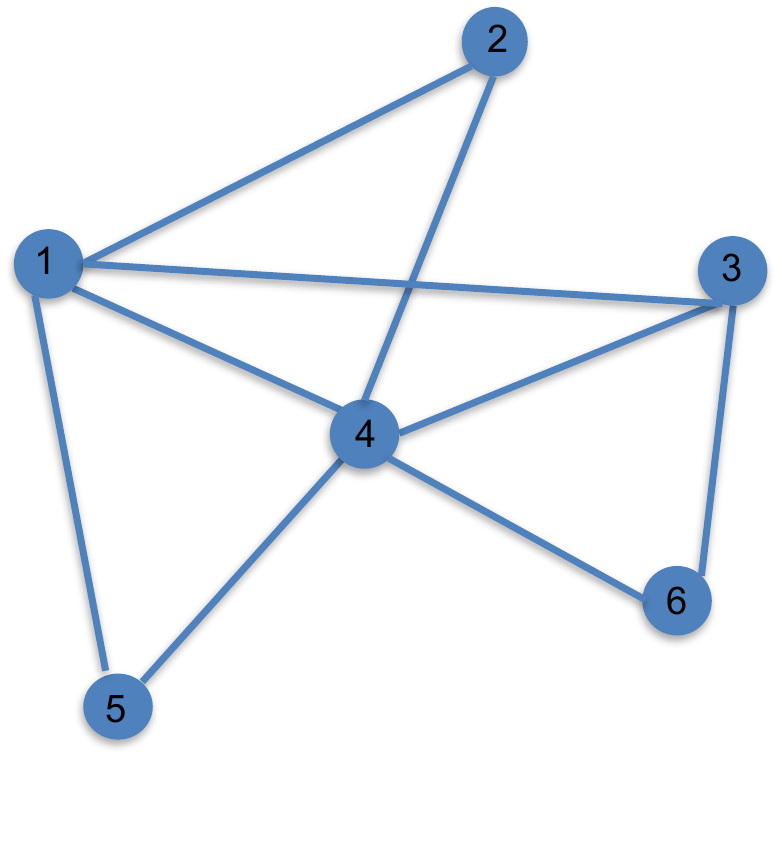}}
\subfigure[Graph with multiple relations]{\label{fig:multi-relation}\includegraphics[scale=0.55]{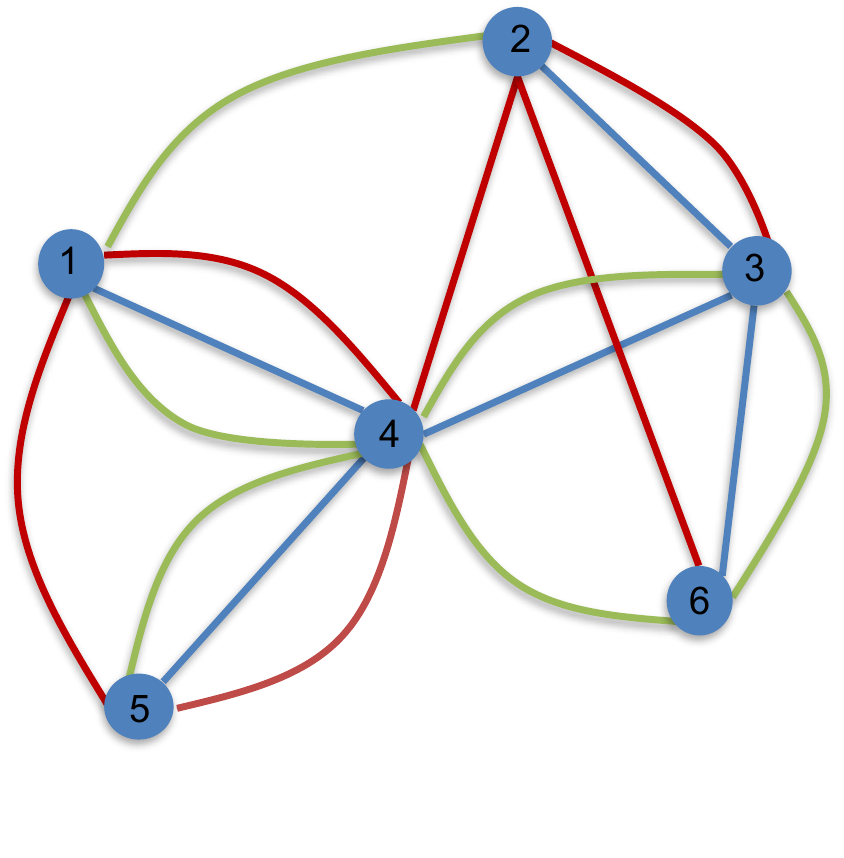}}
\subfigure[Multi-dimensional graph]{\label{fig:multi-dimension}\includegraphics[scale=0.55]{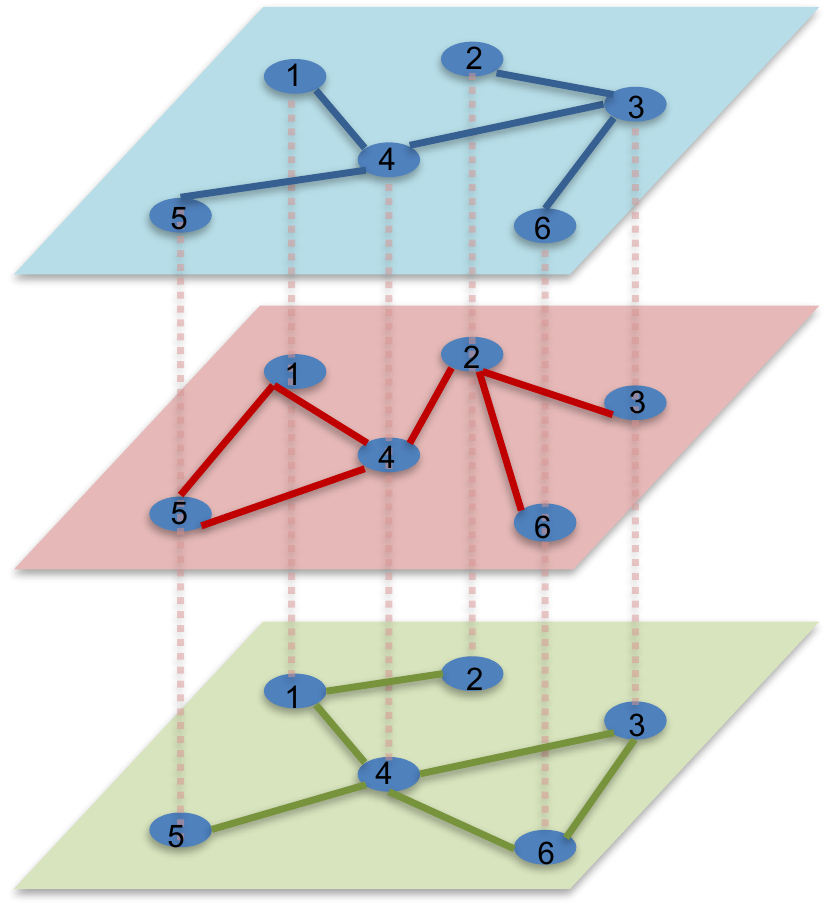}}
\end{center}
\caption{Single-dimensional graph and multi-dimensional graph}
\label{fig:networks}
\end{figure*}

The majority of graph convolutional networks have been developed for single dimensional graphs where only one type of relation can exist between a pair of nodes as shown in Figure~\ref{fig:single-relation}. However, in many real-world graphs,  multiple relations can exist between pairs of nodes simultaneously as shown in Figure~\ref{fig:multi-relation}. For example, in video sharing site like YouTube, users can interact with each other via ``subscription'' relation as well as various social interaction such as ``sharing'' and ``commenting''~\cite{tang-etal2010community}; and in e-commerce platform such as Amazon, users can interact with items via ``view'' and ``purchase'' relations. These complex graphs with multiple relations can be naturally treated as multi-dimensional graphs by viewing each type of relation as a dimension as shown in Figure~\ref{fig:multi-dimension}. Each relation forms a unique graph structure, while the graphs formed by different relations are not independent to each other~\cite{berlingerio-etal2013multidimensional}. Since relations in different dimensions can affect each other, nodes not only can interact with other nodes in each dimension but also can interact with their own ``copies'' in different dimensions~\cite{berlingerio-etal2013multidimensional,hossmann-etal2012collection}. The multiple relations in multi-dimensional graphs bring about additional complexity. It poses tremendous challenges to existing graph convolutional networks, which have been designed for single dimensional graphs. Thus, dedicated efforts are desired to develop graph convolutional networks for multi-dimensional graphs. 

In this work, we study the problem of graph convolutoinal networks for multi-dimensional graphs. In essence, we aim to tackle the following challenges: 1) how to model the node interaction in each dimension while considering the interactions across dimensions; and 2) how to combine the information for multi-dimensional graph convoluitonal networks. These two challenges are addressed by the proposed framework mGCN. Our major contributions can be summarized as follows:
\begin{itemize}
\item We introduce a principled approach to capture the interactions within- and across-dimensions;
\item We propose a multi-dimensional graph convolutional network framework mGCN, which can model rich information in multi-dimensional graphs coherently for representation learning; and
\item We conduct comprehensive experiments on real-world multi-dimensional graphs to demonstrate the effectiveness of the proposed framework. 
\end{itemize}

The rest of this paper is organized as follows. In Section~\ref{sec:model}, we introduce the approach to model the within- and across-dimension interactions and the framework of graph convolutional network for multi-dimensional graphs. In Section~\ref{sec:experiments}, we describe the experimental setting and analyze the experimental results. We then introduce related works in Section~\ref{sec:related_work}. We conclude this work in Section~\ref{sec:conclusions} with possible future directions.

	\section{The Proposed Framework}\label{sec:model}
Before we give details about the proposed framework, we first introduce some notations and definitions we will use in the remaining of this work. Table~\ref{tab:notations} provides a list of major notations we use in the paper. 

\begin{table}[h]
	\begin{center}
		\caption{Notations\label{tab:notations}}
		\begin{tabular}{c|p{6cm}} 
			\hline
			notation & meaning \\ 
			\hline
			$\mathcal{V}$ & the set of nodes in the multi-dimensional graph \\
			\hline
			 $\mathcal{E}_d$ & the set of edges in dimension $d$ of the multi-dimensional graph\\
			\hline
			$N$ & the number of nodes in the multi-dimensional graph\\
			\hline
			$D$ & the number of dimensions in the multi-dimensional graph\\
            \hline
			$N_d(v_i)$ &the set of within-dimension neighbors of node $v_i$ in dimension $d$ \\
            \hline 
            ${\bf A}_d$ &the adjacency matrix of dimension $d$\\
            \hline 

            ${\bf h}_i$ & the general representation of node $v_i$\\
			\hline
                $l$ & the length of the general representation ${\bf h}_i$\\
            \hline
     ${\bf h}_{i,d}$ & the dimension-specific representation of node $v_i$ in dimension $d$\\
         \hline
      $q$ & the length of the dimension-specific representation ${\bf h}_{i,d}$\\

            \hline
            ${\bf H}^k$ & the input general representations fort all nodes in $k$-th layer of mGCN\\
            \hline 
            ${\bf W}_d^k$& the project matrix of dimension $d$ in $k$-th layer of mGCN\\
            \hline
            ${\bf E}^k_d$ & the dimension-specific representations of dimension $d$ for all nodes before aggregation in $k$-th layer of mGCN\\
            \hline
         ${{\bf H}_w}_d^k$ & the results of within-dimension aggregation of dimension $d$ for all nodes in $k$-th layer of mGCN\\
             \hline
             ${{\bf H}_a}_d^k$& the results of across-dimension aggregation of dimension $d$ in $k$-th layer of mGCN\\
             \hline
     		 ${\bf H}^k_d$ & the dimension-specific representations of dimension $d$ for all nodes after aggregation in $k$-th layer of mGCN\\   
             \hline

             ${\bf H}^{k+1}$ &  the output general representations for all nodes in $k$-th layer of mGCN\\
             \hline
             $b_{g,d}$ & the importance score of dimension $g$ to dimension $d$\\
             \hline
		\end{tabular}
	\end{center}

\end{table}

A multi-dimensional graph consists of a set of $N$ nodes $\mathcal{V} = \{v_1,\dots,v_N\}$ and $D$ sets of edges $\{\mathcal{E}_1, \dots ,\mathcal{E}_D\}$, which describe the interaction between nodes in the corresponding $D$ dimensions. These $D$ types of interactions can be also described by $D$ adjacency matrices ${\bf A}_1,\dots, {\bf A}_D$. For a given dimension $d$, its adjacency matrix ${\bf A}_d \in \mathbb{R}^{N\times N}$ describes its connection in this dimension. Let ${\bf A}_d[i,j]$ denote the element in $i$-th row and $j$-th column of ${\bf A}_d$. Then ${\bf A}_d[i,j]=1$ when there is a link between node $v_i$ and $v_j$ in dimension $d$, otherwise $0$.

The graph convolutional network (GCN)~\cite{kipf-etal2016} has been designed for single dimensional graphs. It aims to improve the quality of the representations by aggregating information from neighbor nodes. For a given node, its neighbors are those nodes that are directly connected to it. In other words, traditional GCN utilizes the interactions between nodes and their neighbors to learn node representations.

\begin{figure}[!h]
\centering
\includegraphics[scale = 0.5]{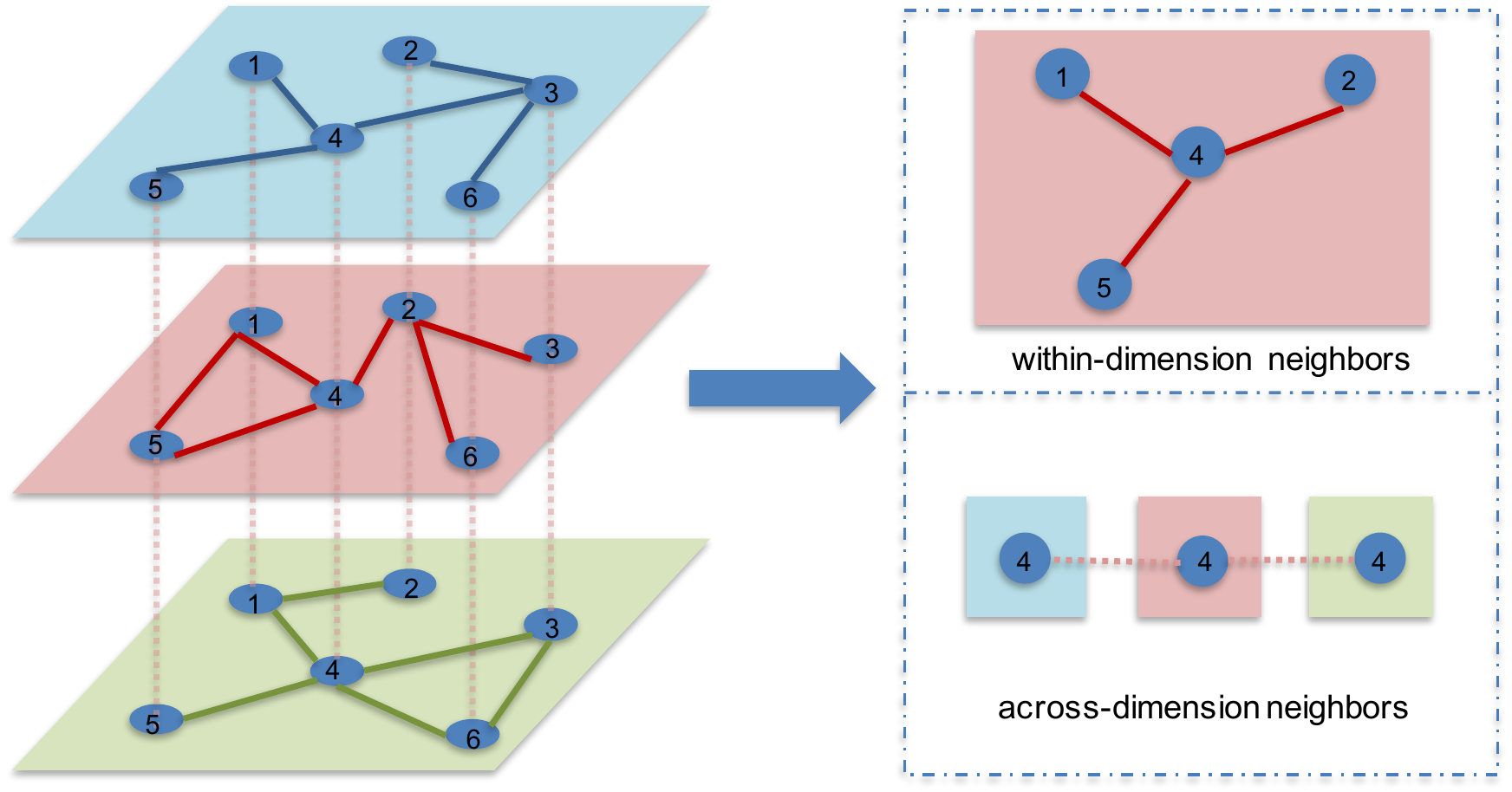}
\caption{There are two types of neighbors in a multi-dimensional graph for a node in a given dimension. For example, for node $4$ in the ``red'' dimension, the within-dimension neighbors are nodes $1$, $2$ and $5$ in the ``red'' dimension, while its across-dimension neighbors are node $4$ in the ``blue'' dimension and node $4$ in the ``green'' dimension.}
\label{fig:multi_dim_gcn}
\end{figure}

However, in a multi-dimensional graph, all the dimensions share the same set of nodes, while nodes interact with each other differently in different dimensions. Different interactions between nodes form varied network structures in different dimensions as shown in Figure~\ref{fig:multi_dim_gcn}. Thus, for a given node, it is likely that it has different neighbors in different dimensions. These different neighbors are specific to each dimension and we call them within-dimension neighbors for each dimension. Furthermore, the same node in different dimensions is inherently related. Thus, for one node in a given dimension, we also need to consider the across-dimension interactions with its own copies in the other dimensions. In this work, we define these copies of a node in the other dimensions as across-dimension neighbors of the given node in the given dimension. More specifically, for a node in a given dimension, there are within-dimension and across-dimension interactions. These two types of interactions lead to two different types of neighbors -- within-dimension and across-dimension neighbors. For a node $v_i$ in a given dimension $d$, the within-dimension neighbors consist of all the nodes that are connected to it in dimension $d$ (or for all $v_j$ where ${\bf A}_d[i,j]=1$). For a node $v_i$ in a given dimension $d$, the across-dimension neighbors consist of its own copies in the other dimensions. For example, in Figure~\ref{fig:multi_dim_gcn}, the within-dimension neighbors for node $4$ in the ``red'' dimension are nodes $1$, $2$ and $5$ in the ``red'' dimension, while its across-dimension neighbors are node $4$ in the ``blue'' dimension and node $4$ in the ``green'' dimension. 


To capture the within-dimension and across-dimension information in each dimension and the general information in the entire multi-dimensional graph, we introduce two representations for each node -- the dimension-specific representations to denote the within-dimension and across-dimension information in each dimension and the general representation to denote the general information in the entire multi-dimensional graph. We first detail these two types of representations, then introduce model components to capture within-dimension and across-dimension information and finally discuss the proposed framework.


\subsection{General and dimension-specific representations}
The dimension-specific representations are corresponding to each dimension, while the general representation is supposed to capture the information from all the dimensions. 
In this section, we introduce the relations between the general representation and dimension-specific representations. More specifically, we describe the procedure of transforming general representation to dimension-specific representations and the procedure of combining dimension-specific representations to form general representation. The procedure not only can help us understand better about the relations between dimension-specific and general representations but also can help us reduce the representation parameters. 

Let ${\bf h}_i \in \mathbb{R}^{l\times 1}$ and ${\bf h}_{i,d} \in \mathbb{R}^{q\times 1}, d=1,\dots, D$ denote the general representation and dimension-specific representations for node $v_i$, respectively. To get the dimension-specific representations ${\bf h}_{i,d}$ from the general representation ${\bf h}_i$, we project the general representation to each dimension. More specifically, we introduce a project matrix ${\bf W}_d\in \mathbb{R}^{q\times l}$ for each dimension $d$ with non-linear activation functions to obtain dimension-specific representations from general representations as follows
\begin{align}
{\bf h}_{i,d} = \text{act}({\bf W}_d \cdot {\bf h}_i)
\label{eq:transform_to_dimension_specific}
\end{align}
where $\text{act}(\cdot)$ is an element-wise non-linear activation function. Modeling the dimension-specific representations as Eq.~(\ref{eq:transform_to_dimension_specific}) can naturally impose the dimension-specific representations of each dimensions to share some information, as they are all generated from the same general representation. Furthermore, we can use the project matrix ${\bf W}_d$ to understand the relations among dimensions. The transformation in Eq.~\eqref{eq:transform_to_dimension_specific} can be also written in the matrix form for all nodes as follows
\begin{align}
{\bf H}_{d} = \text{act}({\bf W}_d \cdot {\bf H})
\label{eq:transform_to_dimension_specific_matrix}
\end{align}
where ${\bf H}_{d} \in \mathbb{R}^{q\times N}$ is the matrix containing the dimension-specific representation for all nodes in dimension $d$, with column $i$ the representation ${\bf h}_{i,d}$ for node $v_i$. Similarly, ${\bf H} \in \mathbb{R}^{l\times N}$ is the matrix containing the general representations for all nodes.

To get the general representation from the dimension-specific representations, we need to integrate the dimension-specific representations. We propose to use a feed forward neural network to perform the combination. We concatenate the $D$ dimension-specific representations as the input of the feed forward network. More specifically, the combination procedure can be represented as 
\begin{align}
	{\bf h}_i = \text{act}({\bf W} \cdot \text{Concat}_{d=1}^D {\bf h}_{i,d}).
	\label{eq:combine}
\end{align}
where ${\bf W }\in{\mathbb{R}^{q\times D\cdot q}}$, Concat() is the function to concatenate the dimension-specific representations and $\text{act}(\cdot)$ is an element-wise non-linear activation function. Similarly, the combination Eq.~\eqref{eq:combine} can also be represented in the matrix form as: 
\begin{align}
	{\bf H} = \text{act}({\bf W} \cdot \text{Concat}_{d=1}^D {\bf H}_{d}).
	\label{eq:combine_matrix}
\end{align}

Similar to traditional GCN, we can have multiple layers in the mGCN. However, to simplify the illustration of the relations of general and dimension-specific representations, we ignore the index of the layer on all parameters in this subsection. For example, the general representation ${\bf h}_i$ in the Eq.~(\ref{eq:transform_to_dimension_specific}) and Eq.~\eqref{eq:combine} can be from different layers. We will add the layer index when we introduce the details of the proposed framework. 

\subsection{Modeling the within-dimension interactions}
In this subsection, we introduce the modeling of the within-dimension interaction. For a node $v_i$, in a given dimension $d$, to capture the within dimension interaction, we perform the single dimensional GCN~\cite{kipf-etal2016} to dimension $d$ based on the within-dimension neighbors. More specifically, it can be represented as
\begin{align}
{{\bf h}_w}_{i,d} = \sum\limits_{v_i\in N_d(v_i)} \hat{\bf A}_d[i,j]\cdot {\bf h}_{j,d}
\label{eq:aggregation_within_dim}
\end{align}
where $N_d(v_i)$ is the set of within-dimension neighbors of node $v_i$ in dimension $d$, $\hat{\bf A}_d = {\bf D}_d^{-1}({\bf A}_d+{\bf I})$ is the row normalized adjacency matrix with self-loop. ${\bf D}_d$ is a diagonal matrix with ${\bf D}_d[i,i]$ the summation of the $i$-th row of $\hat{\bf A} + {\bf I}$. $\hat{\bf A}[i,j]$ is the element of $\hat{\bf A}$ in i-th row and j-th column. Note that we also include the node itself in its within dimension neighbors $N_d(v_i)$. Aggregating information from within-dimension neighbors in Eq.~\eqref{eq:aggregation_within_dim} can also be represented in a matrix form as: 
\begin{align}
{{\bf H}_w}_{d} =  {\bf H}_{d}\cdot \hat{\bf A}_d
\label{eq:aggregation_within_dim_matrix}
\end{align}




\subsection{Modeling the across-dimension interactions}

To model the across-dimension interactions, we perform a similar aggregation as we do for the within dimension interaction but on the across-dimension neighbors. To perform the across-dimension aggregation, we take a weighted average over the dimension-specific representations of node $v_i$: 
\begin{align}
{{\bf h}_{i,d}} = \sum\limits_{g=1,\dots,D} b_{g,d}\cdot {\bf h}_{i,g} 
\label{eq:aggregation_across_dim}
\end{align}
where $\sum\limits_{g=1,\dots,D} b_{g,d}=1$. The weight $b_{g,d}$ models the the importance of dimension $g$ to dimension $d$. Dimensions do not always affect each other equally and it is likely that there are some dimensions that are more similar than others. Naturally, the dimension-specific representation from a more similar dimension should contribute more in the across-dimension aggregation step. However, this kind of correlation information between dimensions is not always explicitly available. Hence, it is difficult for us to get $b_{g,d}$ beforehand. It is desired that these importance scores between dimensions can be learned during the aggregation steps. Recall that dimension-specific representations are obtained by projecting the general representations. The projection matrices are supposed to contain some descriptive information of the dimensions. For example, if two dimensions are highly similar, the two projection matrices should also be highly related. Thus, we introduce an attention mechanism to learn these scores based on these projection matrices. The importance of a dimension $g$ to a dimension $d$ can be learned from the following function:
\begin{align}
p_{g,d} =\text{att}({\bf W}_{g},{\bf W}_{d})
\end{align}
where $\text{att}(\cdot,\cdot)$ is some attention function. The inputs of this attention function are the two projection matrices for the given two dimensions. In this work, we use a bilinear function to model the attention function:
\begin{align}
p_{g,d} = \text{tr}({{\bf W}_{g}}^T {\bf M} {\bf W}_{d})
\end{align}
where $tr()$ is the trace of a matrix and ${\bf M}$ is the parameters to be learned in the bilinear function. We further apply a softmax function to normalize the importance score as:
\begin{align}
b_{g,d} = \frac{\exp(p_{g,d})}{\sum\limits_{g=1}^D \exp(p_{g,d})}
\end{align}

The across-dimension aggregation can also be represented in the matrix form as: 
\begin{align}
{\bf H}_{d} = \sum\limits_{g=1,\dots,D} b_{g,d} \cdot {\bf H}_{g} 
\label{eq:aggregation_across_dim_matrix}
\end{align}
\subsection{Multi-dimensional Graph Convolutional Networks}
\begin{figure*}[!h]
\centering
\includegraphics[scale = 0.55]{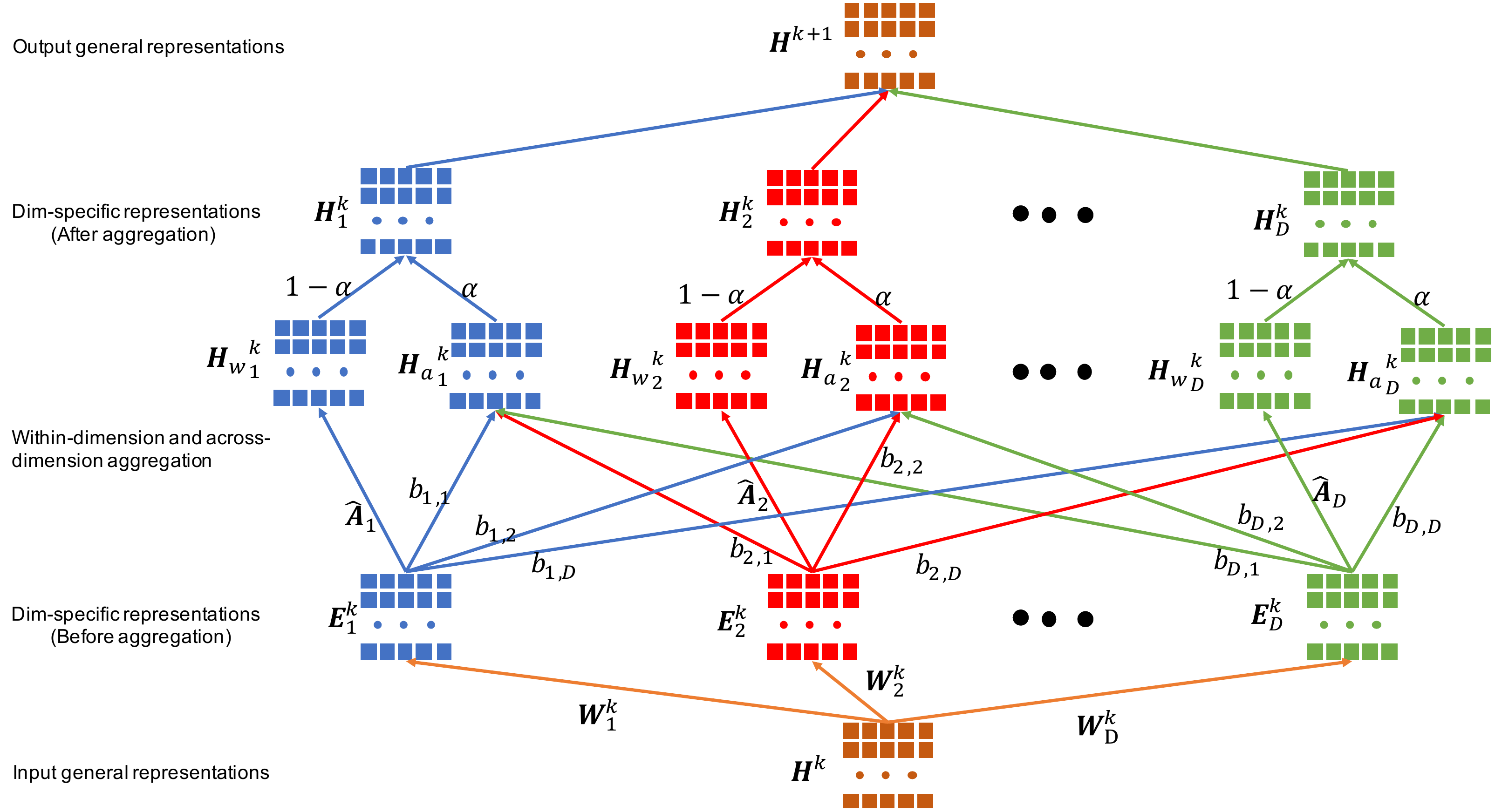}
\caption{The $k$-th layer of mGCN framework}
\label{fig:framework}
\end{figure*}
In this subsection, we introduce the framework of the multi-dimensional GCN, mGCN. Figure~\ref{fig:framework} shows the $k$-th layer of mGCN. The input of this layer is the general representations ${\bf H}^k$ generated in the $(k-1)$-th layer. The general representations ${\bf H}^k$ are then transformed to $D$ dimension-specific representations according to Eq.~\eqref{eq:transform_to_dimension_specific_matrix}
\begin{align}
{\bf E}^k_{d} = \text{act}({\bf W}^k_d \cdot {\bf H}^k),\  d=1,\dots,D.
\label{eq:dim-specific_before_agg}
\end{align}
Here we use ${\bf E}^k_d$ to denote the dimension-specific representations before aggregation to differentiate from the dimension-specific representation after aggregation ${\bf H}^k_d$. 
After the procedures in Eq.~\eqref{eq:dim-specific_before_agg}, we then proceed to the within- and across-dimension aggregation procedures as shown in Figure~\ref{fig:framework} according to Eq.~\eqref{eq:aggregation_within_dim_matrix} and Eq.~\eqref{eq:aggregation_across_dim_matrix}. 
\begin{align}
{{\bf H}_w}^k_{d} & =  {\bf E}^k_{d}\cdot \hat{\bf A}_d,\ d =1,\dots,D.\\
{{\bf H}_{a}}^k_{d}& = \sum\limits_{g=1,\dots,D} b_{g,d} \cdot {\bf E}^k_{g}, \ d = 1,\dots,D.
\end{align}
We then combine the results of the within- and across-dimension aggregations as follows
\begin{align}
{\bf H}^k_d = (1-\alpha)\cdot {{\bf H}_w}^k_{d}  + \alpha \cdot {{\bf H}_{a}}^k_{d},\ d=1,\dots,D,
\end{align}
where $\alpha$ is the hyper-parameter to control the importance between the two components. The combination generates the dimension-specific representations after aggregation, which are denoted as ${\bf H}^k_d,\ d=1,\dots,D$. Finally, we combine these dimension-specific representations to get new general representations ${\bf H}^{k+1}$ according to Eq.~\eqref{eq:combine_matrix}:
\begin{align}
	{\bf H}^{k+1} = \text{act}({\bf W^k} \cdot \text{Concat}_{d=1}^D {\bf H^k}_{d}).
\end{align}
We introduce the $k$-th layer of the multi-dimensional GCN and the output is the new general representations ${\bf H}^{k+1}$, which can serve as the input of the ($k+1$)-th layer. In each layer, we use the output of the previous layer as the input. To initialize the procedure,  the input general representations ${\bf H}^0$ is needed. These initial general representations could be features associated with the nodes, representations learned by some network embedding methods, or even randomly initialized representations. Let the input be ${\bf X}$, then, we initialize ${\bf H}^0 = {\bf X}$. The output of the multi-dimensional GCN is the general representations ${\bf H^{K}}$ formed in the ($K-1$)-th layer. For convenience, we denote ${\bf Z}:= {\bf H}^{K}$. 

The parameters in the model include the projection matrices ${\bf W^k_d},\ d=1,\dots,D$, ${\bf W}^k$ of the fully connected combination layer and ${\bf M}^k$ of the attention function for $k=0,\dots,K-1$. To train the model, different loss functions can be designed. For example, we could use supervised information from the node labels. In this work, we design an unsupervised loss function using the linkage information. More specifically, we model the probability that a link existing between node $v_i$ and node $v_j$ in dimension $d$ as 
\begin{align}
p(1|v_i,v_j,d) = \sigma(({\bf W}_d^{K+1}\cdot {\bf z}_i)^T({\bf W}_d^{K+1}\cdot {\bf z}_j));
\end{align}
where ${\bf W}_d^{K+1}$ is the projection matrix for dimension $d$, ${\bf z}_i$ is the $i$-th column of ${\bf Z}$ and $\sigma(\cdot)$ is the sigmoid function $\sigma(x) = 1/(1+\exp(-x))$. The probability that there is no link between node $v_i$ and node $v_j$ in dimension $d$ can be modeled as
\begin{align}
p(0|v_i,v_j,d) &= 1- p(1|v_i,v_j,d) \nonumber\\
& = 1 - \sigma(({\bf W}_d^{K+1}\cdot {\bf z}_i)^T({\bf W}_d^{K+1}\cdot {\bf z}_j))\nonumber \\
& =  \sigma(-({\bf W}_d^{K+1}\cdot {\bf z}_i)^T({\bf W}_d^{K+1}\cdot {\bf z}_j))
\end{align}
We can now model the negative logarithm likelihood as 
\begin{align}
L = &-\log \left( \prod\limits_{(v_i,v_j,d) \in \mathcal{T}_p} p(1|v_i,v_j,d)\cdot \prod\limits_{(v_i,v_j,d) \in \mathcal{T}_n} p(0|v_i,v_j,d) \right) \nonumber\\
= &-\sum\limits_{(v_i,v_j,d) \in \mathcal{T}_p} \log p(1|v_i,v_j,d) \nonumber\\
&-\sum\limits_{(v_i,v_j,d) \in \mathcal{T}_n} \log p(0|v_i,v_j,d),
\label{eq:loss}
\end{align}
where $\mathcal{T}_p$ is the set of positive samples, which consists of all the existing links in the multi-dimensional graphs. These links can be denoted as triplets $(v_i,v_j,d)$, which means node $v_i$ and node $v_j$ are connected in dimension $d$. $\mathcal{T}_n$ is the set of negative samples, which consists of all the triplets $(v_i,v_j,d)$ with no link between $v_i$ and $v_j$ in dimension $d$. 

\subsection{An Optimization Method}
Real-world graphs such as social networks or computer networks are usually very sparse~\cite{barabasi2016network}. The number of existing links compared to the number of missing links is small and usually it can be linearly bounded by the number of nodes~\cite{barabasi2016network}. This means the size of $\mathcal{T}_n$ is large. It is computationally expensive to consider all the pairs in the loss Eq.~\eqref{eq:loss}. To solve this issue, we use the negative sampling approach proposed in~\cite{mikolov-etal2013distributed}. For each $(v_i,v_j,d)\in \mathcal{T}_p$, we fix $v_i$ and $d$ and randomly sample $n$ nodes that are not connected to node $v_i$ in dimension $d$. These $n$ samples are put into the set of the negative samples. In this way, the size of $\mathcal{T}_n$ is only $n$ times as large as the size of $\mathcal{T}_p$. 

For large graphs with millions of nodes, it is prohibited to perform the within-dimension aggregation in Eq.~\eqref{eq:aggregation_within_dim_matrix} for all nodes at the same time. Furthermore, if we adopt a mini-batch procedure to optimize the loss function, it is not necessary to calculate the representations for all the nodes during each mini-batch. Only those nodes that are within $K$-hops of the nodes in the mini-batch training samples are involved. Therefore, in practice we only calculate the representations for those involved nodes. However, if we use all the neighbors for given nodes (k-hop neighbors, for $k=1,\dots,K$) in the within-dimension aggregation step, the number of nodes involved can still get very large, in the worst case, it can still involve all the nodes in the graph. Hence, we decide to sample $s$ neighbors for a given node $v_i$ from its within-dimension neighbors when performing the within-dimension aggregation as similar in~\cite{hamilton-etal2017inductive}. We adopt a mini-batch ADAM~\cite{kingma-etal2014adam} to optimize the framework. 
	\section{Experiments}\label{sec:experiments}
In this section, we validate the effectiveness of the proposed framework by conducting the link prediction and node classification tasks on two real-world multi-dimensional graphs. We first introduce the two multi-dimensional graphs we use in this paper. Then, we describe the link prediction and node classification tasks with discussions of the experimental results. Further experiments are conducted to understand the importance of within- and across- dimension aggregation step in mGCN.
\subsection{Datasets}

\begin{table}[h]
	\begin{center}
		\caption{Statistics of datasets\label{tab:statistics}}
		\begin{tabular}{c |c|c } 
			\hline
			 & DBLP& Epinions \\ 
			\hline
			number of nodes &138,072 &15,108\\
			\hline
			number of  edges &2,015,650 & 485,154\\
			\hline
			number of dimensions & 20&5\\
			\hline
			number of labels & 10 & 15\\
			\hline 
		\end{tabular}
	\end{center}

\end{table}

In this subsection, we introduce the two datasets we use in this paper. Some important statistics are shown in Table~\ref{tab:statistics}. Detailed descriptions of the two datasets are as follows: 

\begin{itemize}
\item \textbf{DBLP:} DBLP\footnote{http://dblp.uni-trier.de/} is a computer science bibliography website. It holds bibliographic information on major computer science journals and proceedings. We collect all publication records from major computer science journals and conferences during $1998$-$2017$. We then treat the co-authorship in each year as different relations and form a co-authorship multi-dimensional graph. In this multi-dimensional graph, each year forms a dimension. If two authors cooperate to publish a paper in a given year, we create a link between them in the dimension of the given year. We also assign labels to the authors according to all the papers they published. More specifically, each paper belongs to a high-level category such as ``Theoretical Computer Science'' and ``Machine Learning''. We assign the category which most of the author's papers belong to as his/her label. 
\item \textbf{Epinions} Epinions\footnote{http://www.epinions.com/} is a general review site, where users can write reviews for products and rate helpfulness for reviews written by other users. Users in this site can also form trust and distrust relations. We form a $5$-dimensional graph based on $5$ different relations between users: 1) co-review: if two users review some common products, we create a link between them in the co-review dimension; 2) helpfulness-rating: we create a link between two users in the helpfulness-rating dimension if a user rates the reviews written by the other user; 3) co-rating: if two users rate some common reviews, we create a link between them in the co-rating dimension; 4) trust: We create the trust dimension based on the trust relation between users; and 5) distrust: We create the distrust dimension based on the distrust relation between users. We also assign labels to the users according to the products they reviewed. More specifically, each product belongs to a category. We assign the category which most of the products reviewed by the user belong to as his/her label. 

\subsection{Comparison algorithms}
Our method learns representations for each node in the multi-dimensional graph. So we compare our method with representative single dimensional and multi-dimensional representation learning algorithms. To apply single-dimensional algorithms, we aggregate multi-dimensional graph into a single dimensional graph. Note that we can also apply single-dimensional algorithms on each dimension and then aggregate the representations. We do not include this strategy in this work as it is likely to suffer from the data sparsity problem. Next, we describe these methods as below: 
\begin{itemize}
\item Non-negative matrix factorization (NMF)~\cite{lin2007projected}. We apply NMF to the adjacency matrix of the aggregated single dimensional graph and use the factorized matrix as the representations. 
\item LINE~\cite{tang-etal2015line}. LINE is a recent proposed network embedding method, which can learn representations for all the nodes in the  graph. It can only be applied to single dimensional graph. Thus, as similar in NMF, we aggregate the multi-dimensional graph as a single dimension graph and apply LINE on it. We use the default setting of LINE while setting the number of training samples to $800$ millions. 
\item node2vec~\cite{grover-etal2016node2vec}. node2vec is a state-of-the-art network embedding method. It can only be applied to single dimensional graph. Thus we apply node2vec on the aggregated single-dimensional graph. We follow the default setting of node2vec and conduct a grid search for $q$ and $p$ as in~\cite{grover-etal2016node2vec}. 
\item MINES~\cite{ma-etal2018multi}. MINES is a recent proposed network embedding method designed for multi-dimensional network. We apply MINES to the multi-dimensional graph to learn the node representations.

\item GCN. This is a variant of the traditional GCN~\cite{kipf-etal2016}. The difference is that we adopt the same loss as in mGCN instead of the original semi-supervised loss. Additionally, we also sample neighbors instead of using all the neighbors. This method is designed for single dimensional graphs, so we apply it on the single dimensional graph aggregated from the multi-dimensional graph.

\item mGCN-noa. This is a variant of our method mGCN, where the weights $b_{g,d}$ in the across-dimension aggregation Eq.~\eqref{eq:aggregation_across_dim_matrix} are all $\frac{1}{D}$, which means we simply take the average over all the dimension specific representations when perform the within-dimension aggregation. 
\end{itemize}

\subsection{Node Classification}
In the node classification task, we try to predict labels for unlabeled nodes. In the experiments setting, we hide the labels for a fraction of nodes and use the label information of the labeled nodes and the graph structure to perform the node classification. 

As in~\cite{perozzi-etal2014deepwalk}, we try different fractions of nodes to hide their labels, the remaining nodes with labels are treated as training samples to train a classifier. In this paper, we set the ratio of training samples to $10\% - 90\%$ with a step-size of $20\%$. We use the node representations as the input features of nodes and train a logistic regression classifier. The nodes with label hided are treated as the testing set. For each of the setting, we try $10$ different splits of the training and testing and report the average performance of the $10$ experiments. The metrics we use to measure the performance are $F_1$-macro and $F_1$-micro score as in~\cite{grover-etal2016node2vec,perozzi-etal2014deepwalk}. For all the methods, we set the length of the representation to $64$ for fair comparison. For node classification task, we use the final general representations ${\bf Z}$ for our mGCN and mGCN-noa. We use the representation generated by LINE as input for Epinions dataset for GCN, mGCN and mGCN-noa. For DBLP, we use the representation learned by node2vec as the input for these methods. We set the value of $\alpha$ to $0.5$, the number of negative samples $n$ to $2$, the number of sample neighbors $s$ to $10$ and the number of layers $K$ to $1$. More layers can be stacked and we leave this as a possible future direction. 

\subsubsection{Experiments Results}
The results of node classification for Epinions and DBLP dataset are shown in Figure~\ref{fig:ep_results} and Figure~\ref{fig:dblp_results}, respectively. Note that the performance of NMF on both datasets are not comparable to the other methods. More specifically, the representations learned by NMF only achieve $F_1$-macro $0.2380$, $F_1$-micro $0.4012$ on the Epinions dataset and $F_1$-macro $0.2380$, $F_1$-micro $0.3580$ on the DBLP dataset both under the $90\%$ training setting. Thus, we do not include the performance of NMF in the figures. We can make the following observations from these results:
\begin{itemize}
\item Our method mGCN outperforms all the baselines under all the settings on both datasets, which shows the effectiveness of our method.
\item The performance of mGCN is better than GCN, which suggests that utilizing the multi-dimensional relations in the multi-dimensional graph is necessary and our proposed method mGCN can facilitate them well to help learn better representations. 
\item mGCN is better than MINES, which shows the effectiveness of mGCN to capture the within- and across-dimension information in a better way. 
\item mGCN is better than mGCN-noa, which indicates the effectiveness of the attention mechanism in the across-dimension aggregation step. 
\end{itemize}

\begin{figure*}
\begin{center}
\subfigure[Epinions: $F_1$-macro]{\label{fig:ep_f1_macro}\includegraphics[scale=0.4]{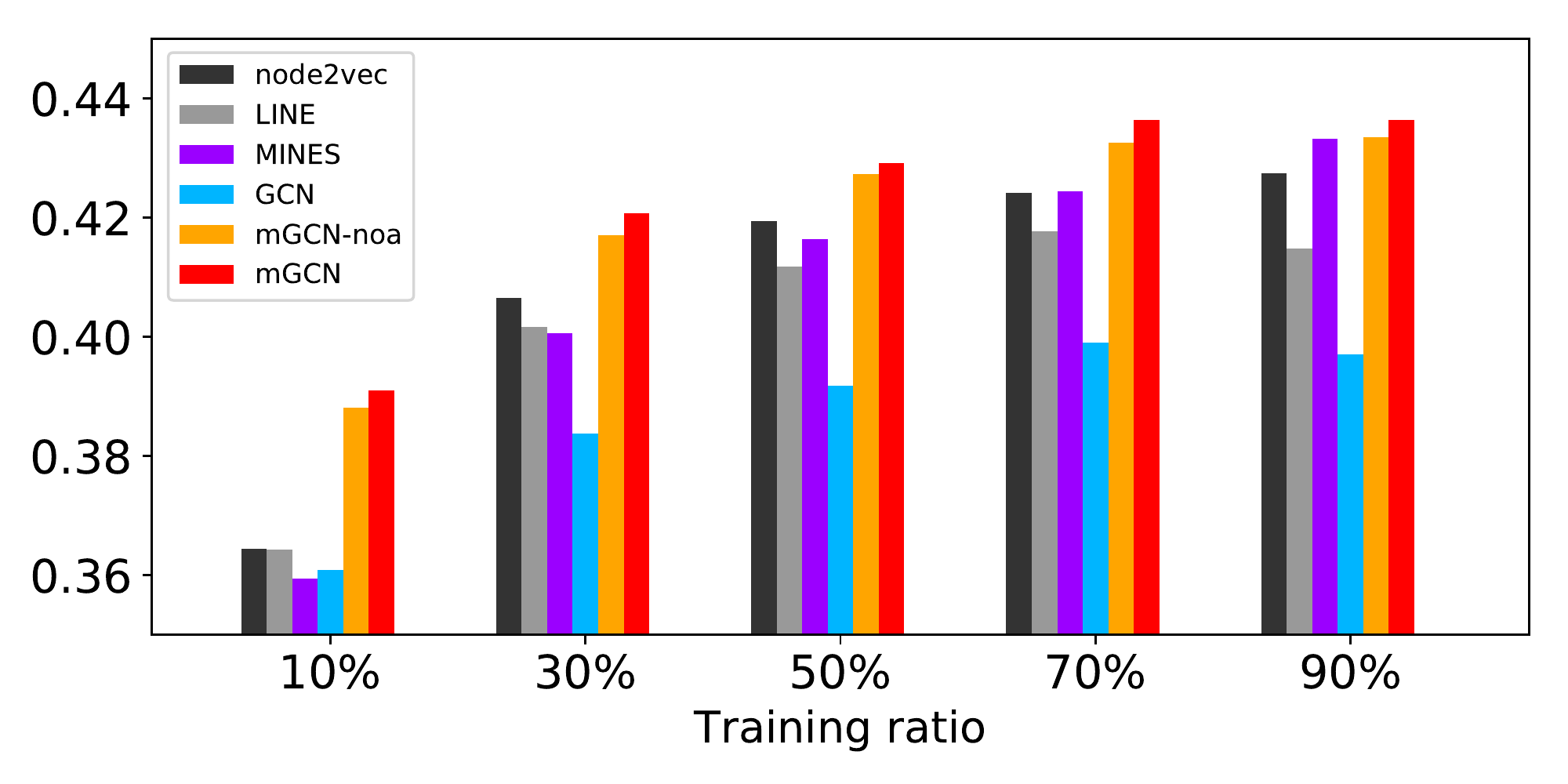}}
\subfigure[Epinions: $F_1$-micro]{\label{fig:ep_f1_micro}\includegraphics[scale=0.4]{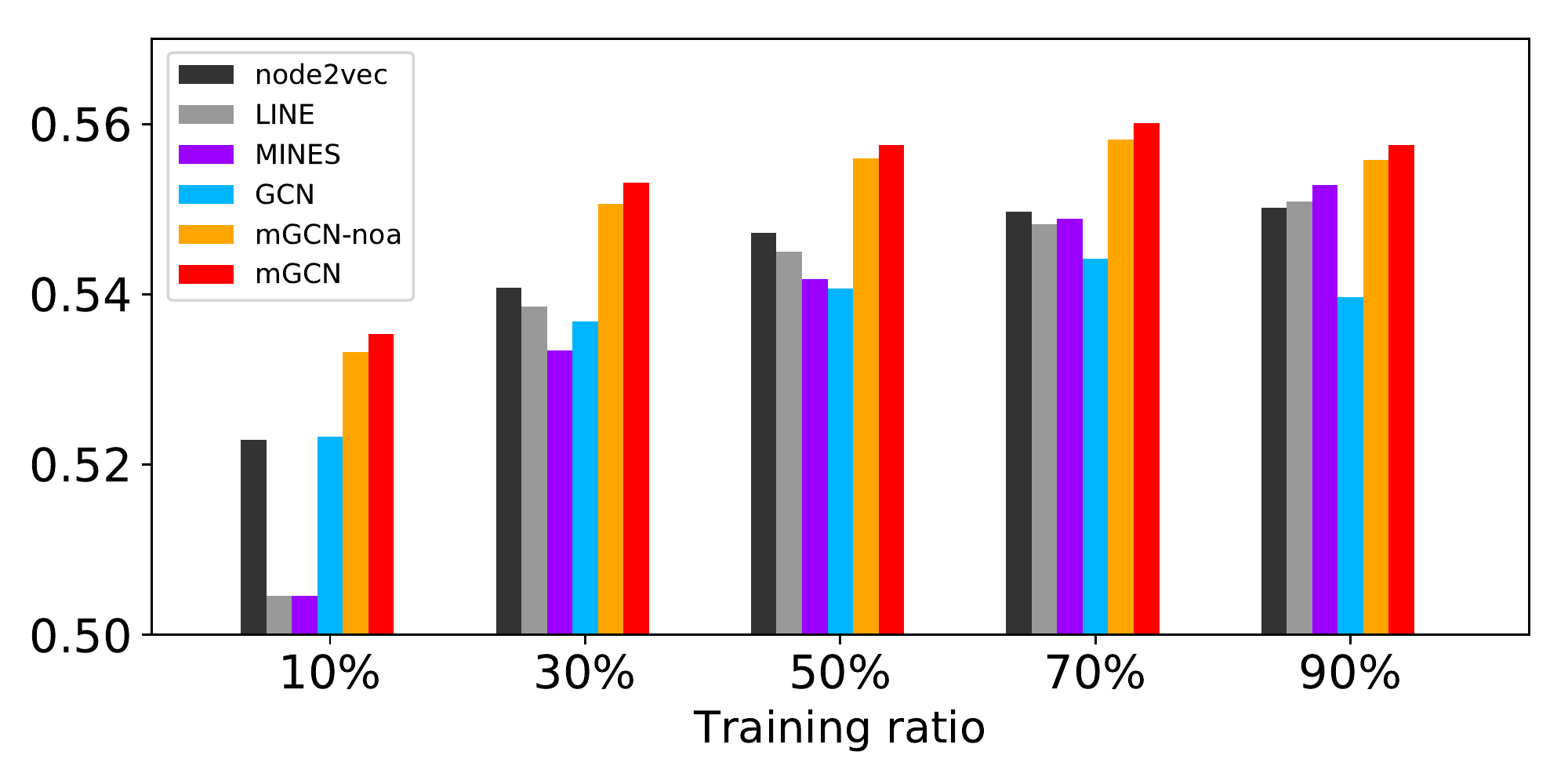}}
\end{center}
\caption{Performance Comparison of Node classification on Epinions dataset}
\label{fig:ep_results}
\end{figure*}


\begin{figure*}
\begin{center}
\subfigure[DBLP: $F_1$-macro]{\label{fig:dblp_f1_macro}\includegraphics[scale=0.4]{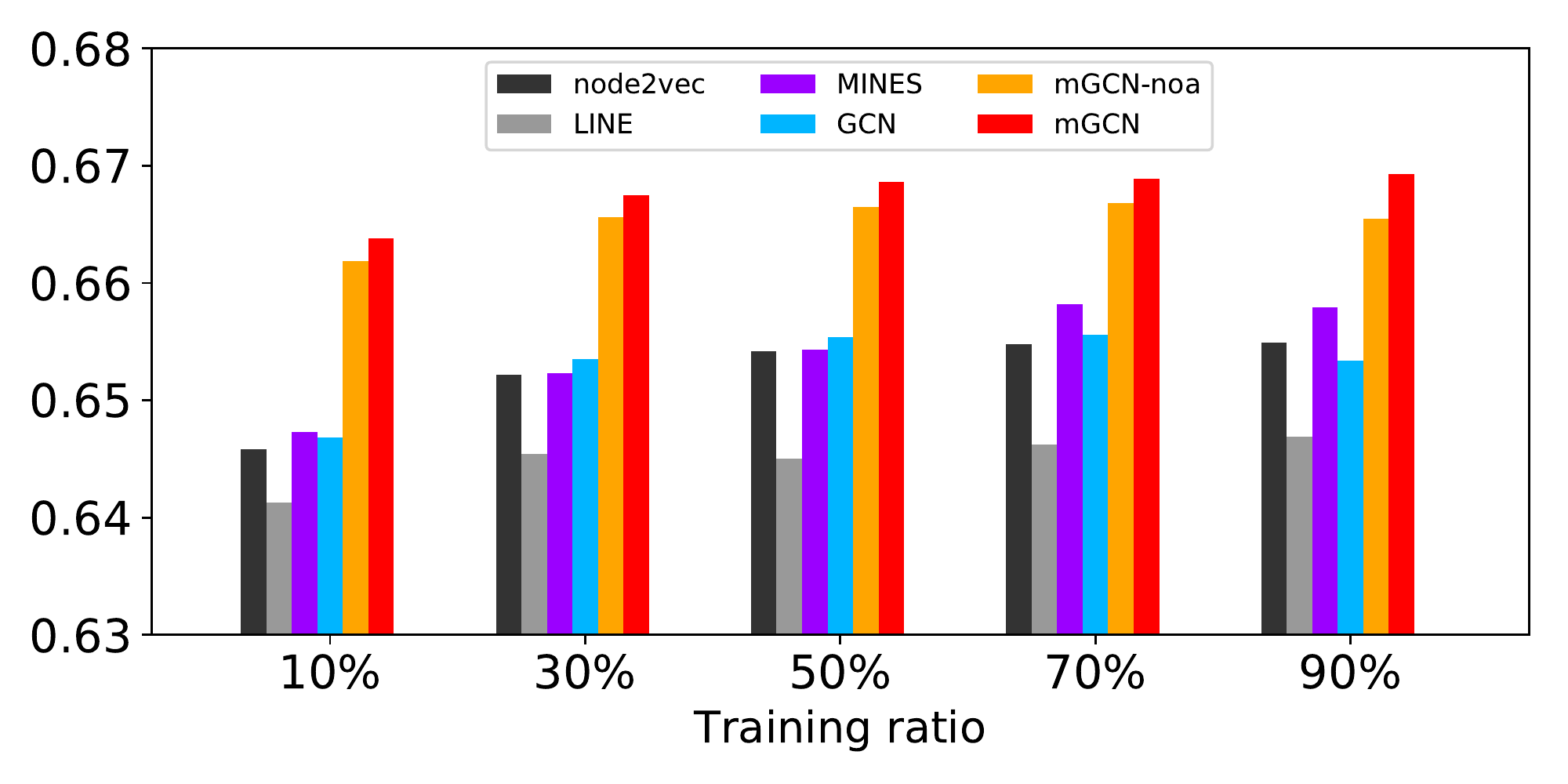}}
\subfigure[DBLP: $F_1$-micro]{\label{fig:dblp_f1_micro}\includegraphics[scale=0.4]{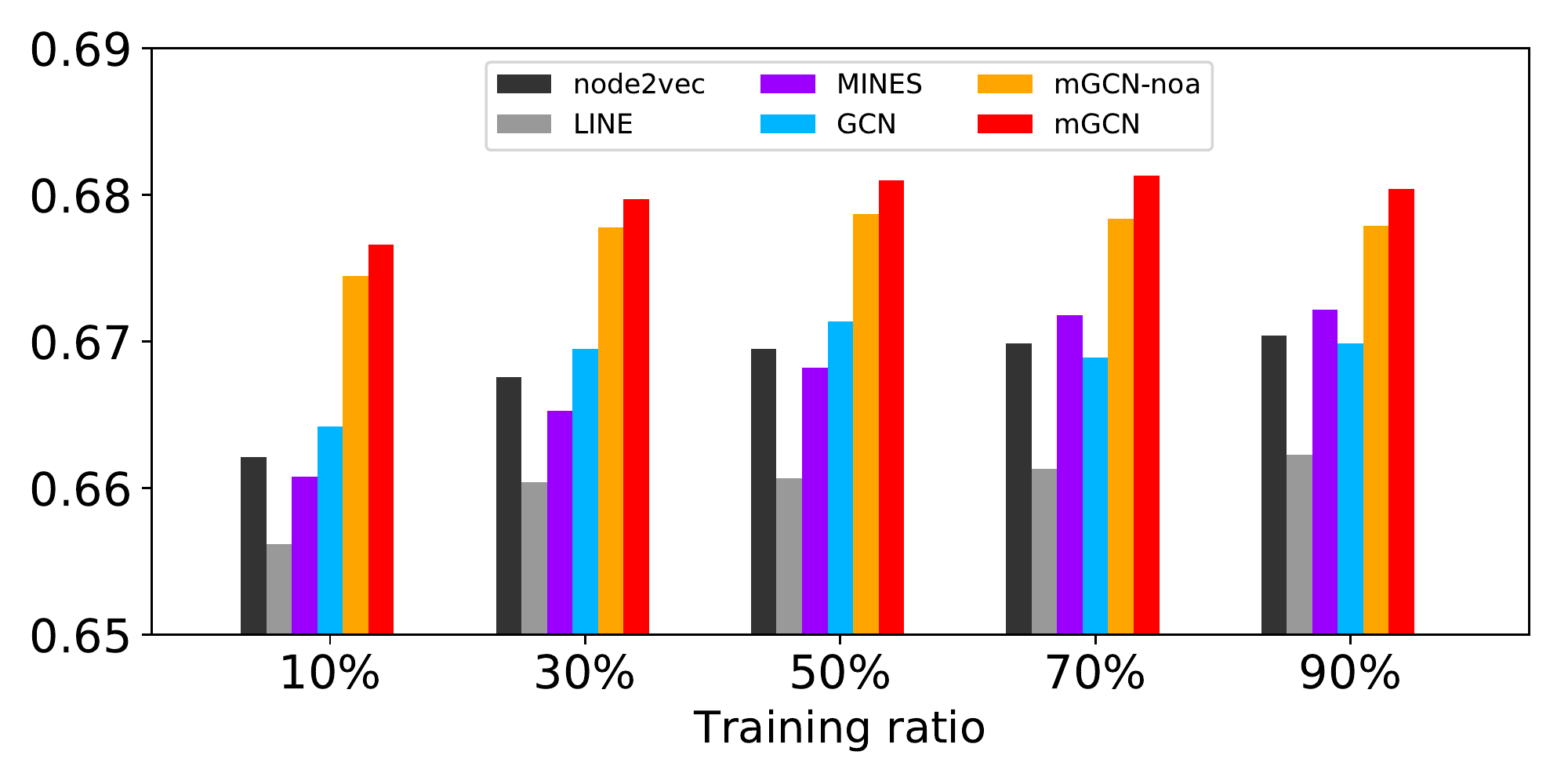}}
\end{center}
\caption{Performance Comparison of Node classification on DBLP dataset}
\label{fig:dblp_results}
\end{figure*}

\subsection{Link prediction}
In the link prediction task, we try to predict whether a non-existing link will emerge in the future based on the current graph. In the traditional setting of link prediction task for single dimensional graph, a fraction of links are removed from the graph and used as the ground truth and then we try to predict their existence using the remaining graph. In the multi-dimensional graph setting, we perform link prediction in different dimensions, separately. When performing link prediction in dimension $d$, we first remove a fraction of links from dimension $d$. Then, for each removed link, if the two nodes connected by this link are connected in the other dimensions, we also remove those links in the other dimensions. After removing these links, we use the remained graph to learn the representations for  all the methods. We then formulate the link prediction task as a binary classification problem as in~\cite{grover-etal2016node2vec} using the combination of the features of node pairs as the input features. Different combination operations can be used to get the feature for a node pair from the features of the two nodes. In this paper, we use the element-wise multiplication, as it achieves the best performance among all the operations in~\cite{grover-etal2016node2vec}. When a pair of nodes are connected by a link in dimension $d$, this pair of nodes is labeled as $1$, otherwise $0$. To form the training set of the binary classifier, we add all the links in dimension $d$ in the remaining graph as positive samples and randomly generate equal number of non-connected node pairs as negative samples. We form the testing set in a similar way, using the removed links as positive samples and randomly generated equal number of negative samples. For the DBLP data, we only perform link prediction task on the last dimension (Year 2017), as it is not reasonable to use future information to predict past links. We conduct link prediction under two different settings, one with $50\%$ links removed and the other one with $70\%$ links removed. For the Epinions dataset, we perform link prediction tasks in  each of the $5$ dimension with $20\%$ of the links removed. We use $AUC$ score as the metric to measure the performance of link prediction as in~\cite{grover-etal2016node2vec}.

For the experiments on both dataset, we set the length of the representations to $64$ for fair comparison. For mGCN and mGCN-noa, we use the ${\bf W}^{K+1}_d\cdot {\bf Z}$ as the representations when conducting link prediction on dimension $d$. As similar in the node classification task, we use the representations learned by LINE and node2vec as input for Epinions and DBLP dataset respectively. We set the value of $\alpha$ to $0.5$, the number of negative samples $n$ to $2$, the number of sample neighbors $s$ to $10$ and the number of layers $K$ to $1$.


\subsubsection{Experiments Results}

The results for link prediction on the Epinions dataset are shown in Table~\ref{tab:ep_lp_results}. We conduct link prediction experiments in each dimension of the Epinions data the set. The $AUC$ of each dimension is reported in Table~\ref{tab:ep_lp_results}, where dim $0$-$4$ denote the co-review dimension, the helpfulness-rating dimension, the co-rating dimension, the trust dimension and the distrust dimension, respectively. The average performance over all dimensions is reported in the last column. We can make the following observations from this table. 
\begin{itemize}
\item The link prediction performance on different dimensions varies a lot, which indicates that the network structure are indeed different in each dimension. 
\item The performance of the methods designed for single dimensional graph (LINE, node2vec, etal) is worse than MINEs, mGCN-noa and mGCN. This suggests that simply ignoring the different types of the relations and combining the multi-dimensional graph as single dimensional graph may cause loss of information. mGCN performs better than GCN, which further indicates that the mGCN model effectively captures the unique information from multi-dimensional graph. 
\item  On average, mGCN outperforms mGCN-noa. This suggests that our attention mechanism can capture the relations between dimensions well. 
\end{itemize}

\begin{table}[!h]
\caption{Performance comparison of link prediction in terms of AUC on Epinions dataset}
\begin{center}
	\begin{tabular}{ |c|c|c|c|c|c| c|} 
		\hline
		method & dim 0 & dim 1 & dim 2 & dim 3 & dim 4&average \\ 
		\hline
		NMF&0.9463&0.8092&0.9381&0.8909&0.9066&0.8982\\
		\hline		
		LINE&0.8612&0.7747&0.8100&0.8088&0.8888&0.8287 \\
		\hline
		node2vec&0.8713&0.7866&0.8773&0.8121&0.8732&0.8441\\
		\hline
		MINES&{0.9621}& {0.8268}&0.9572 & 0.8036&0.8711&0.8842\\
		\hline
		GCN & 0.9080&0.7512& 0.8226&0.8194&0.9002& 0.8403\\
        \hline
        mGCN-noa & 0.9349 & 0.8031&0.9125&0.8642&0.9654 & 0.8960\\
        \hline 
        mGCN & 0.9434& 0.8214&0.9273&0.8795 & 0.9648 & 0.9072\\
        \hline


	\end{tabular}
\end{center}

\label{tab:ep_lp_results}
\end{table}

\begin{figure}[!h]
	\centering
	\includegraphics[scale=0.4]{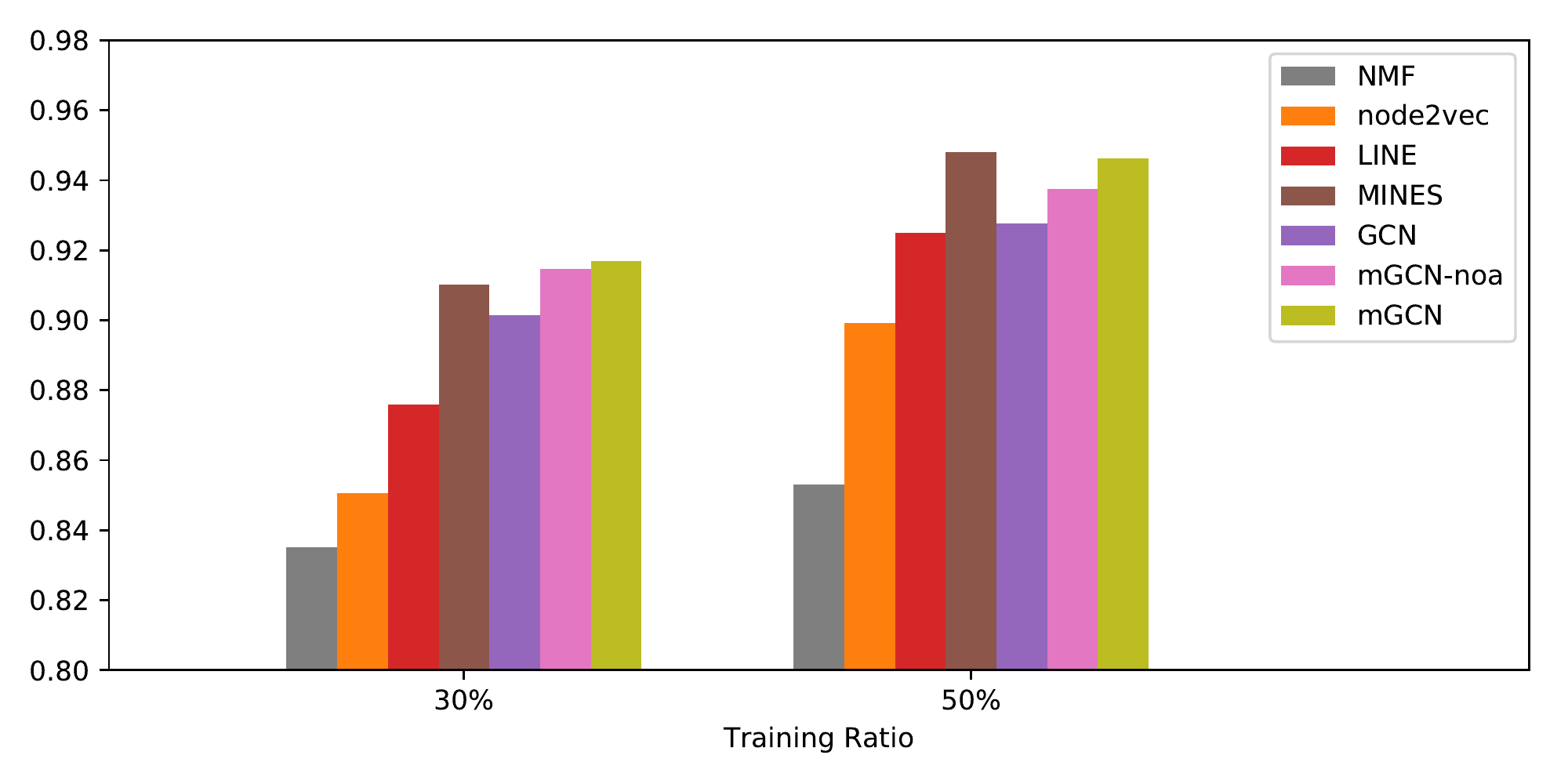}
	\caption{ Performance comparison of link prediction in terms of AUC on the DBLP dataset}
	\label{fig:lp_dblp}
\end{figure}

The link prediction performance on DBLP dataset are shown in Figure~\ref{fig:lp_dblp}. We make similar observations as those on the Epinions dataset. 

\subsection{Components Analysis}
In this subsection, we analyze how the two components, i.e., the within-dimension aggregation and across-dimension aggregation, affect the performance of the mGCN model. Their contributions to the proposed framework are controlled by $\alpha$. Thus to answer the question, we set $\alpha$ to $5$ different values $0,0.3,0.5,0.7,1$. We  only show the node classification experiments on the representations learned by mGCN with these different values of $\alpha$ since we have similar observations to other settings. The length of the representations is set to $64$. 


\begin{figure}
\begin{center}
\subfigure[DBLP: F1-macro]{\label{fig:pa_dblp_macro}\includegraphics[scale=0.3]{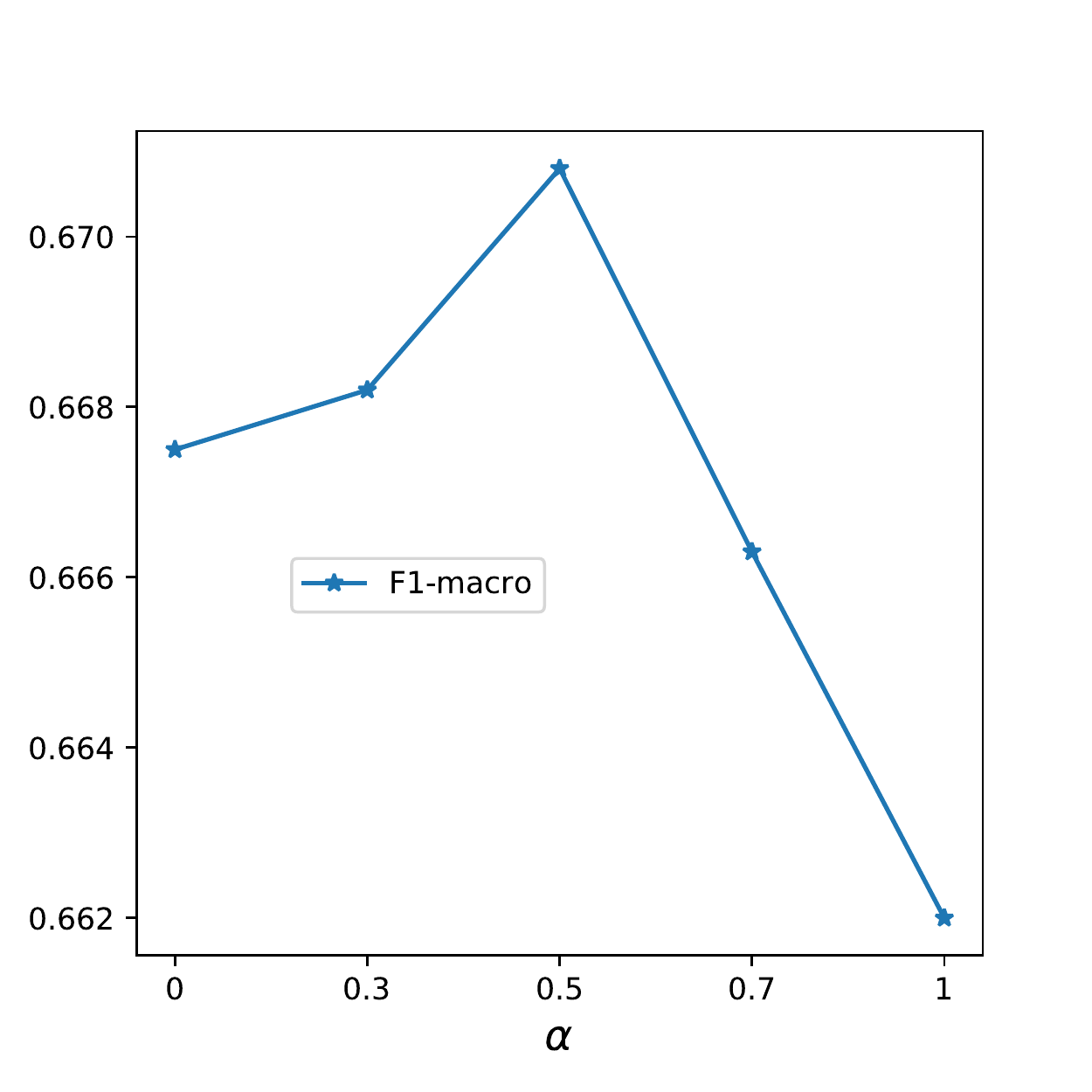}}
\subfigure[DBLP: F1-micro]{\label{fig:pa_dblp_micro}\includegraphics[scale=0.3]{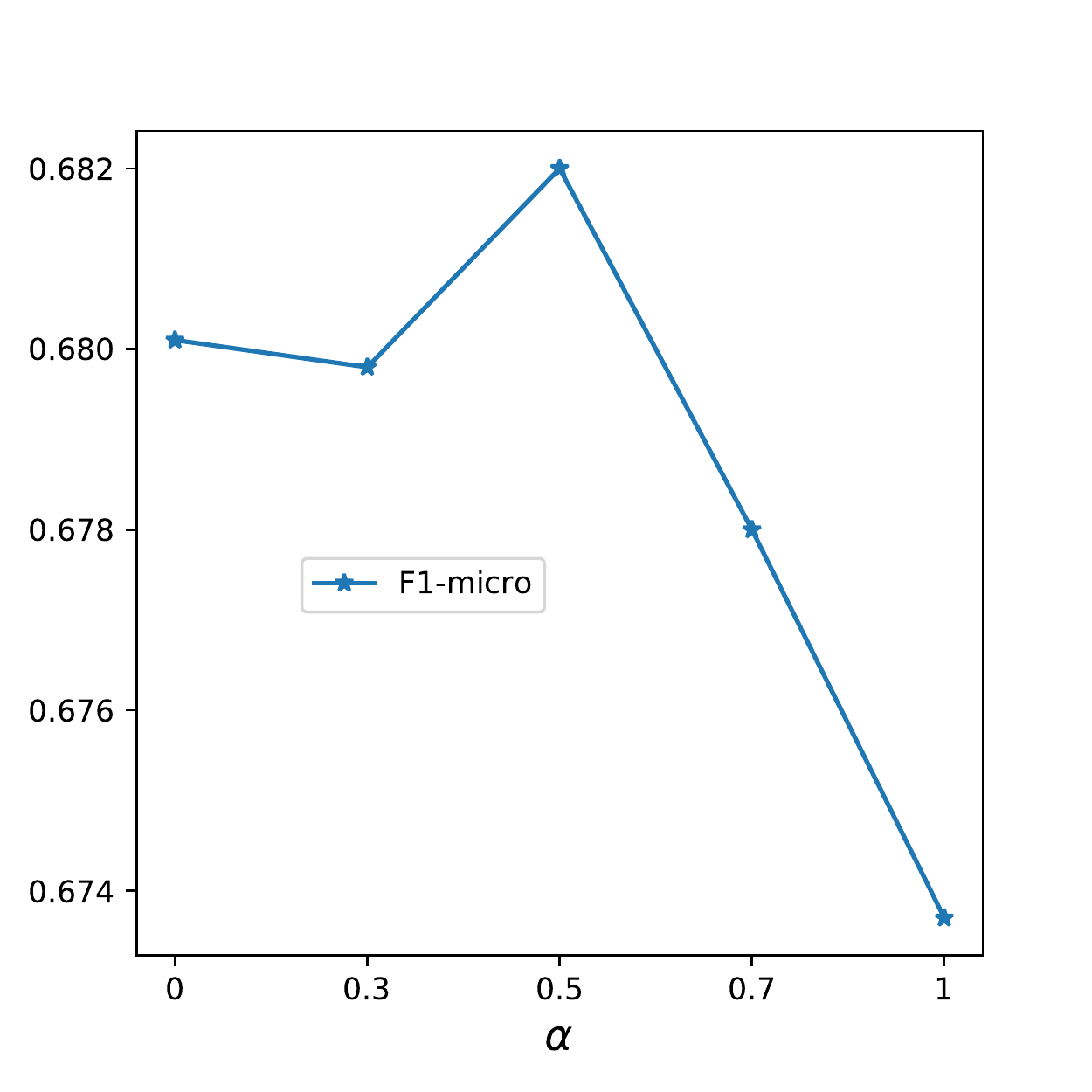}}
\end{center}
\caption{The performance of m-GCN with different $\alpha$}
\label{fig:pa_dblp}
\end{figure}
\end{itemize}

The performances of mGCN with different values of $\alpha$ are shown in Figure~\ref{fig:pa_dblp} in terms of $F_1$-macro and $F_1$-micro score. As we can observe from the figure, the $F_1$-macro score first increases as the value of $\alpha$ gets larger, then it decreases after the value of $\alpha$ is larger than $0.5$. A similar trend can be found for the $F_1$-micro score. The change of the performance with the different $\alpha$ values indicates that both components are important for the model. For example, the model cannot perform well when either one is missing ($\alpha=0$ or $\alpha=1$). It also indicates that the information in the within-dimension and across-dimension are complementary to each other.

	\section{Related work}\label{sec:related_work}
Learning appropriate representations for nodes in graph is essential for many graph related machine learning and data mining tasks. Early methods such as Isomap~\cite{tenenbaum-etal2000global} and Laplacian Eigenmap~\cite{belkin-etal2002laplacian} try to perform dimension reduction on the graph. These methods usually involve computational expensive eigen-decomposition and thus are not scalable when the graph is large. Recent methods inspired by word embedding learning such as DeepWalk~\cite{perozzi-etal2014deepwalk}, LINE~\cite{tang-etal2015line} and node2vec~\cite{grover-etal2016node2vec} can scale to large graph with millions of nodes. Recent efforts~\cite{levy-etal2014neural,qiu-etal2018network} have been made to connect these methods with matrix factorization. There are also works~\cite{wang-etal2016structural,chang-etal2015heterogeneous} trying to use deep learning techniques to learn node representations. Two recent surveys provide comprehensive overviews on graph representation learning algorithms~\cite{cai-etal2017comprehensive,hamilton-etal2017representation}. Recently, there are also some network embedding methods~\cite{ma-etal2018multi,qu-etal2017attention} designed for multi-dimensional graphs. These methods are also inspired by the word embedding learning, but extended to the multi-dimensional graphs.

Convolutional Neural Networks (CNNs)~\cite{lecun-etal1998gradient} break-throughly improve the performance in images, videos related task~\cite{krizhevsky-etal2012imagenet,lawrence-etal1997face,karpathy2014large}. This shows its great power to learn good representations on regular grid data. However, graph or network data are highly irregular. Efforts have been made to generalize CNNs to graphs~\cite{bruna-etal2013spectral,henaff-etal2015deep,duvenaud-etal2015convolutional,li-etal2015gated,defferrard-etal2016convolutional,kipf-etal2016,hamilton-etal2017inductive,schlichtkrull-etal2017modeling}. Some of these methods~\cite{bruna-etal2013spectral,henaff-etal2015deep,duvenaud-etal2015convolutional,li-etal2015gated,defferrard-etal2016convolutional} focus on generalizing CNNs for graph level-representation learning. The other methods~\cite{kipf-etal2016,hamilton-etal2017inductive,schlichtkrull-etal2017modeling} works on learning node representations using convolutional neural networks. Most of the aforementioned GCN methods are designed for single dimensional graphs. However, many real-world graphs are multi-dimensional with multiple types of relations. Some examples and basic properties of multi-dimensional graph can be found in~\cite{boccaletti-etal2014structure,berlingerio-etal2013multidimensional}. In this work, we propose to study graph convolutional neural network for multi-dimensional graphs.

	\section{Conclusions and Future Work}\label{sec:conclusions}
In this paper, we develop a novel graph convolutional network for multi-dimensional graphs.  We propose to use dimension-specific representations to capture the information for node in each dimension  and general representations to capture the information for node over the entire graph. Particularly, we propose to capture the information from both within-dimension and across-dimension interactions when modeling dimension-specific representations in each dimension. We then use a fully connected layer to combine these dimension-specific representations to the general representations. We conduct comprehensive experiments on two real-world multi-dimensional networks. The experimental results demonstrate the effectiveness of the proposed framework. 

In this work, we take a weighted average to combine the representations from the within-dimension aggregation and across-dimension aggregation. More advanced combination method such as non-linear function or even feed-forward neural networks can be tried. Many real-world graphs or networks are naturally evolving. Thus it would be an interesting topic to consider the temporal information when modeling graph convolutional networks. For example, a possible direction is to consider the created timestamp of link when modeling the within-dimension aggregation. 
	\balance
	\bibliographystyle{ACM-Reference-Format}
	\bibliography{yaoma.bib} 
	
\end{document}